\documentclass[a4paper,UKenglish,cleveref,autoref,thm-restate]{lipics-v2021}

\title{Homotopy Type Theory in Isabelle} 

\titlerunning{Homotopy Type Theory in Isabelle} 

\author{Joshua Chen}{University of Nottingham, United Kingdom \and \url{https://joshchen.io}}{joshua.chen@nottingham.ac.uk}{https://orcid.org/0000-0001-5041-0794}{}

\authorrunning{J. Chen} 

\Copyright{Joshua Chen} 



\category{Short paper} 

\relatedversion{} 

\supplement{Source code available at \url{https://github.com/jaycech3n/Isabelle-HoTT/tree/ITP2021}.}

\funding{}

\acknowledgements{The author thanks the reviewers of this paper, as well as those for an earlier draft, for their helpful comments and suggestions for improvement.}

\nolinenumbers 

\hideLIPIcs  

\EventEditors{Liron Cohen and Cezary Kaliszyk}
\EventNoEds{2}
\EventLongTitle{12th International Conference on Interactive Theorem Proving (ITP 2021)}
\EventShortTitle{ITP 2021}
\EventAcronym{ITP}
\EventYear{2021}
\EventDate{June 29--July 1, 2021}
\EventLocation{Rome, Italy (Virtual Conference)}
\EventLogo{}
\SeriesVolume{193}
\ArticleNo{16}

\usepackage{stmaryrd, mathtools, ebproof, booktabs}
\newcommand{\ccolon}{::}
\newcommand{\imp}{\Longrightarrow}
\newcommand{\all}{\bigwedge}
\newcommand{\eq}{\equiv}
\newcommand{\hastype}{\mathtt{has\_type}}

\newcommand{\Ol}{\mathtt{O}}
\newcommand{\Sl}{\mathtt{S}}

\newcommand{\U}{\mathtt{U}}

\newcommand{\SigInd}{\mathtt{SigInd}}

\newcommand{\enc}{\mathbf{\varepsilon}}

\newcommand{\code}[1]{\texttt{#1}}
\newcommand{\meta}{\texttt{?}}
\newcommand{\iarg}{\texttt{\{\}}}

\begin{document}

\maketitle

\begin{abstract}
This paper introduces Isabelle/HoTT, the first development of homotopy type theory in the Isabelle proof assistant.
Building on earlier work by Paulson, I use Isabelle's existing logical framework infrastructure to implement essential automation, such as type checking and term elaboration, that is usually handled on the source code level of dependently typed systems.
I also integrate the propositions-as-types paradigm with the declarative Isar proof language, providing an alternative to the tactic-based proofs of Coq and the proof terms of Agda.
The infrastructure developed is then used to formalize foundational results from the Homotopy Type Theory book.
\end{abstract}

\section{Introduction}

Isabelle \cite{isabelle2020} is a simply typed proof assistant and logical framework.
Of its multiple object logics Isabelle/HOL \cite{npw20} is arguably the best known, however many other logics have been created since Isabelle’s inception and are still bundled along with its distribution.
Among these early logics is Isabelle/CTT (\emph{constructive type theory}) \cite{pau19}, which is based on extensional Martin-L\"{o}f type theory but which has not been further developed.
In light of considerable recent progress in the field of dependent type theory, it seems appropriate to revive support for this in Isabelle.
This paper aims to do this by introducing \emph{Isabelle/HoTT}, the first development of homotopy type theory in Isabelle.

Widely accepted folklore in the theorem proving community holds that a sufficiently strong logical framework can in principle be used to encode and work in any foundational theory of equal or lesser strength.
However, a drawback to this approach
is that one then has to implement the foundation-specific infrastructure on one's own, while working within the additional constraints imposed by the framework.
This becomes particularly clear when the formalism of the framework logic is sufficiently different from that of the object logic, as in our current case.

The Isabelle/HoTT project may thus be viewed in three distinct but related ways:
\begin{itemize}
\item As the beginnings of an Isabelle formalization of homotopy type theory.
\item As a dependently typed Isabelle object logic which improves on Isabelle/CTT with necessary supporting infrastructure for type checking, term elaboration, proof term abstraction and tactics.
\item As a practical case study on the implementation issues discussed in the previous paragraph.
\end{itemize}

This is a short paper on ongoing work.
Isabelle/HoTT currently lacks automation for function definitions, datatypes, and advanced features like higher inductive types (although these may be manually defined or postulated).
Despite this, it is already able to formalize nontrivial results from the Homotopy Type Theory book \cite{ufp13}.
In addition, although the logic presented here is formulated in the axiomatic style of the HoTT book, one could use the same approach to develop two-level type theory \cite{ack16, acks19} and cubical type theory \cite{ahh18, cchm18} in Isabelle.




Isabelle/HoTT is implemented as a library of Standard ML and Isabelle theory files.
References to specific files are given as footnotes throughout this paper, and the source code is available online at \url{https://github.com/jaycech3n/Isabelle-HoTT/tree/ITP2021}.

\paragraph*{Related Work}
One of the earliest object logics for Isabelle was Paulson's Isabelle/CTT \cite{pau19} for constructive type theory with extensional equality.
Indeed, the fundamental ideas of using resolution to perform type checking and inference, and of discharging subgoals in order of increasing flexibility, already appear here.
Isabelle/HoTT improves on this work by implementing universes, an intensional equality type, as well as better integration of type inference and implicit elaboration into the proof process.

Another recent study in developing homotopy type theory in a logical framework appears in work by Barras and Maestracci \cite{bm21}, where they present a partial embedding of de Morgan cubical type theory \cite{cchm18} using rewrite rules in the $\lambda\Pi$-calculus modulo logic of Dedukti \cite{abc+16}.
Our encoding of axiomatic HoTT in simple type theory is more straightforward, allowing us to focus instead on issues arising from integrating the simply and dependently typed paradigms of the meta and object logics.

The largest computer developments of homotopy type theory are well known and use the Coq and Agda proof assistants \cite{bgl+17, hott-agda}.
In these settings the theory is developed synthetically, and in the case of Coq the source code was directly modified in order to implement new features required by the theory.
In our case the trusted prover code is untouched, and we simply extend Isabelle/Pure with new features using its existing logical framework facilities.

\section{Logical Foundations} \label{sec:logical-foundations}

\subparagraph{Judgments.}
We begin\footnote{\code{mltt/core/MLTT.thy}}, as usual, by declaring a meta type $o$ of terms of the object logic, and a constructor $\hastype \ccolon o \to o \to prop$ (written as usual with an infix colon) to encode the typing assertion.
Since we implement Russell-style universes, types are themselves terms and must have the same meta type, and in this way we will effectively have a set of untyped terms in higher order abstract syntax.
Working with Tarski-style universes would allow us to maintain the syntactic type/term distinction with separate meta types, at the cost of having to introduce interpretation operators everywhere.

Judgmental equality of the type theory is shallowly embedded using the Isabelle/Pure equality $\eq$.
This forgets type information, but allows us to more easily reuse the simplifier to compute terms.

\subparagraph{Universes.}
We postulate a set of levels isomorphic to the standard natural numbers with their usual order, by declaring a meta type $lvl$ and constants $\Ol$, $\Sl$ and $<$.
Universes are formed by a constructor $\U \ccolon lvl \to o$, and we axiomatize rules governing the ordering of levels, as well as the hierarchy and cumulativity of universes.

\subparagraph{Types and Terms.}
The constants for formers, constructors and eliminators for the $\Pi$, $\Sigma$ and identity types are postulated using Church-style typing.
Type families as well as function arguments to dependent eliminators are encoded using meta instead of object lambda terms.
For example, in theoretical presentations the $\Sigma$-eliminator might be given by a term $\SigInd(A, B, C, f)$ whose third and fourth arguments are, respectively, a type family $C \colon (\Sigma\,A\,B) \to U$ and a function $f \colon \Pi (x \colon A) (y \colon B(x)).\,C(x, y)$ defining the value of $C$ on all pairs $(a, b) \colon \Sigma\,A\,B$.
In the encoding, these are instead given as the simply typed meta functions $C \ccolon o \to o$ and $f \ccolon o \to o \to o$.
However, after the $\Pi$-type has been encoded, Isabelle's implicit coercion mechanism (Section 12.3 of \cite{wen20}) is used to coerce object functions into meta functions, which allows users to ignore this distinction most of the time.

\subparagraph{Inference Rules.}
Following Jacobs and Melham \cite{jm93}, we define an encoding $\enc$ from the judgments of dependent type theory into Isabelle/Pure by sending
$x_1 \colon A_1, \dotsc, x_n \colon A_n \vdash \mathcal{I}$
to the universally-quantified implication
\[ \all x_1, \dotsc, x_n.\ x_1 \colon A_1 \imp \dotsb \imp x_n \colon A_n \imp \enc(\mathcal{I}), \]
where $\enc(t \colon T) \vcentcolon= t \colon T$ and $\enc(a \eq b \colon T) \vcentcolon= a \eq b$ are the encodings of typing and equality discussed above.
This encoding is recursively extended to inference rules by defining
\[
    \enc\left(
        \begin{prooftree}
            \hypo{\mathcal{J}_1} \hypo{\cdots} \hypo{\mathcal{J}_k}
            \infer3[]{\mathcal{J}}
        \end{prooftree}
    \right)
    \vcentcolon=
    \big(
        \enc(\mathcal{J}_1) \imp \dotsb \imp \enc(\mathcal{J}_k) \imp \enc(\mathcal{J})
    \big).
\]
Note that entailment and derivability are both translated to Pure implication.
The usual rules for formation, introduction, elimination, computation and congruence of $\Pi$, $\Sigma$ and equality types (see e.g.\ \cite{ufp13}) can then be axiomatized.

More generally, a statement
\begin{equation}
\all \vec{x}.\ P_1(\vec{x}) \imp \dotsb \imp P_k(\vec{x}) \imp Q(\vec{x}) \label{eq:statement}
\end{equation}
in Isabelle/HoTT may be viewed as an extended form of type-theoretic judgment
\[ \Gamma \vdash t \colon T \quad \text{or} \quad
\Gamma \vdash t \eq s \colon T \ \text{(where $T$ is implicit)}, \]
where contexts $\Gamma = \big( P_1(\vec{x}), \dotsc, P_k(\vec{x}) \big)$ are also allowed to contain equality judgments, and where the metavariables $\vec{x} = \{x_1, \dotsc, x_m\}$ may appear throughout.
In particular, the $x_i$ need not be explicitly typed by the context $\Gamma$.
Such occurrences typically appear as unification variables in type checking and elaboration problems.

\section{Proof Infrastructure} \label{sec:proof-infrastructure}




\subparagraph{Implicits and Elaboration.}
Implicit arguments and term elaboration are crucial to working in a dependently typed system.
We declare constants \meta{} and \iarg{} representing, respectively, holes and implicit arguments, together with a theorem attribute \code{implicit} and an Isabelle syntax phase operation \code{make\_holes}.\footnote{\code{mltt/core/implicits.ML}}
We can then use the usual definitional facilities together with the \code{implicit} attribute in the usual manner, e.g.
\begin{lstlisting}[mathescape]
definition Id_i (infix "=" 110) where [implicit]: "x = y $\eq$ x ={} y"
\end{lstlisting}
where \code{x =A y} is the fully explicit notation for the equality type.
The implicits \iarg{} in such definitions will be parsed by \code{make\_holes} into holes \meta{}, which are then further converted into schematic variables (i.e.\ Isabelle/Pure metavariables) in goal statements.

The implementation of implicit arguments as schematic variables means that a general goal statement in Isabelle/HoTT is schematic.
Such goals are not very well supported by the existing Isar commands, so we define new goal keywords \code{Lemma}, \code{Theorem}, etc.\ (replacing \code{lemma}, \code{theorem} etc.)\footnote{\code{mltt/core/goals.ML}}\ as well as a command \code{assuming} (replacing \code{assume}).\footnote{\code{mltt/core/elaborated\_statement.ML}}
These call the type checker on assumptions to infer their implicit arguments and thus instantiate all metavariables before passing them to the regular context assumption mechanism.\footnote{\code{mltt/core/elaboration.ML}}

\subparagraph{Proof Terms.}
Consider the task of automatically abstracting proof terms into definitions.
A theorem stated in a dependently typed system is given by a single type \`{a} la Curry-Howard.
In contrast, in the LCF-style setting of Isabelle the assumptions of a theorem statement are typically available as facts in an Isabelle/Isar proof context, which are lifted to premises after the conclusion has been proved.
In particular, these premises are \emph{not} bound by the type of the theorem's conclusion.
This distinction is exactly the isomorphism---given by the $\Pi$-introduction rule---between open terms with variables typed by a nonempty (type-theoretic) context, and closed terms of a $\Pi$-type (i.e.\ lambda terms).

Hence, in Isabelle, the proof term in a theorem's conclusion must be abstracted over all variables typed by the premises, in order to form a meta lambda term.
This is then wrapped up into a definition.
This functionality is available as a modifier \code{(def)} to the goal statement keywords discussed previously.

\subparagraph{Induction/Elimination Rules.}
In dependent type theory, given a predicate $C \colon A \to U$, the elimination rule for $A$ is used to prove $C(a)$ for all $a \colon A$.
Crucially, this requires that $C$ encodes all the assumptions needed to prove its conclusion.
As previously noted, in Isabelle these assumptions may instead appear out in the Isar context, and thus in order to be able to apply elimination rules correctly we must ensure that such ``free-floating'' assumptions are pushed into the type of the goal.

Concretely, this involves checking the conclusion $Q$ of a goal \eqref{eq:statement} for variables that are typed by premises or Isar context facts $P_i$, and then using $\Pi$-formation to push these assumptions into the object logic predicate $C$.
Furthermore, since the Isar context is unordered and there may be typing dependencies among these assumptions, we first topologically sort them by $\lesssim$, where $t_i \colon T_i \lesssim t_j \colon T_j$ if $t_i$ is a subterm of $T_j$.
This process is automated by infrastructure introducing the \code{elim} attribute and proof method.\footnote{\code{mltt/core/elimination.ML}, \code{mltt/core/tactics.ML}}

\subparagraph{Propositional Equality and Calculational Reasoning.}
The identity type $x =_A y$ is presented as an inductive family over its endpoints $x$, $y$.
Its induction principle is subject to the same general considerations for elimination rules discussed above, and the proof method \code{eq} for reasoning with path induction is essentially a special case of the \code{elim} method.

Rewriting (aka \emph{transport}) along propositional equalities is given by a method \code{rewr}.\footnote{\code{hott/Identity.thy}}
We additionally adapt Isar's calculational reasoning (Sections 1.2 and 2.2.4 of \cite{wen20}) to so-called \emph{calculational} types, which are types $T \colon A \to A \to U$ that have a notion of composition $\diamond \colon \Pi\{x, y, z \colon A\}.\ T(x, y)\to T(y,z) \to T(x,z)$ expressing transitivity of $T$.
After declaring a calculational type and its transitivity rule with the \code{calc} keyword\footnote{\code{mltt/core/calc.ML}} and \code{trans} attribute we can then use the familiar idiom ``\code{have\textellipsis{}also have\textellipsis{}finally show}'' to construct transitive chains in proofs.
By design, this technology is also general enough to support reasoning with chains of homotopies $f \sim g$.

\section{Type Checking} \label{sec:typechecking}

The type checker\footnote{\code{mltt/core/types.ML}} is a key component integrated throughout the infrastructure described above.
It is used by goal commands to perform implicit elaboration, hooked in to proof methods to automatically discharge ancillary typing conditions that arise throughout the course of a proof, and installed as an Isabelle simp-solver\footnote{An Isabelle simplifier component that solves subgoals arising from conditional simplification rules.} to enable typed term reduction.
It is also available as a standalone method \code{typechk}.

At its core is a tactic that recursively resolves goals against the type inference (i.e.\ formation, introduction and elimination) rules, suitable facts from the local Isar context, any additional rules declared with the \code{type} attribute, and the conversion rule.
It is restricted to judgments $t \colon T$ where $t$ is rigid (i.e.\ where the head of $t$ is a constant) and $T$ may be schematic.
Combined with unification this yields a bidirectional type checking/inference algorithm, which is syntax-directed on the collection of type inference rules since every rule in this collection types a term with unique head.
Nondeterminism is introduced when resolving against context facts and rules from the user-modifiable \code{type} theorem collection, and here backtracking allows the checker to try all possible options.
If it fails to completely solve an inference problem, the type checker will return the goals on which it failed to the user for further refinement.

The conversion rule
$\all a\,A\,A^\prime.\ a \colon A \imp A \eq A^\prime \imp a \colon A^\prime$
introduces normalization into the type checker; the simplifier is used to solve the second proof obligation.
This is currently somewhat rudimentary since definitional unfolding is not yet implemented, but this is expected to be relatively straightforward to add.

As already noted by Paulson, the order in which subgoals are tackled in a type inference problem matters greatly, as the large number of metavariables---especially with implicit arguments---creates potentially many unification candidates and too large a search space if not resolved against the correct rule.
He mitigates this by using a filter-and-repeat technique to attempt the subgoals with the fewest metavariables first; we achieve a similar effect by carefully ordering the premises of inference rules according to the criteria for bidirectional type systems set out by Dunfield and Krishnaswami \cite{dk20}.

\section{Formalization}

\begin{figure}
\centering
\fbox{
    \includegraphics[width=0.9\textwidth]{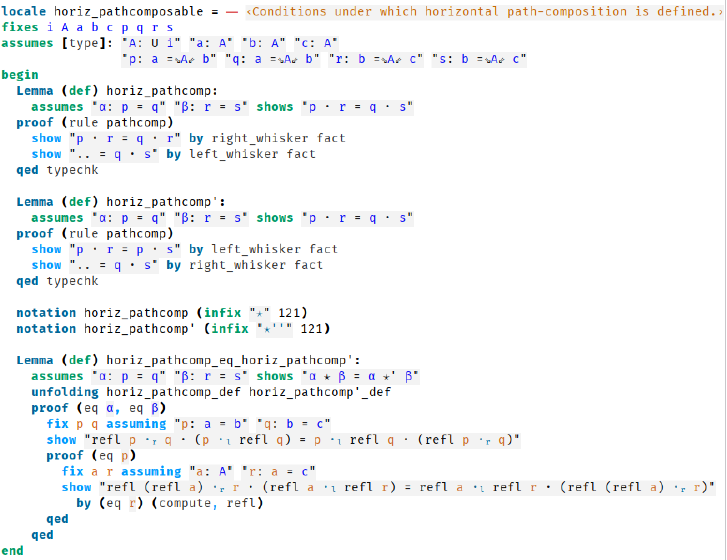}
}
\caption{Example: Horizontal composition.} \label{fig:example}
\end{figure}

The object logic developed is used to formalize material from the first chapters of the Homotopy Type Theory book in Isabelle2020 \cite{isabelle2020}, including results on equality, homotopies and equivalences, and more.\footnote{\code{hott/*.thy}}
\cref{fig:example} shows an example proof that the two ways one can define horizontal composition of equalities on a type $A$ are equal\footnote{\code{hott/Identity.thy}.}, which is an intermediate result en route to the proof of the Eckmann-Hilton argument for $\Omega^2(A)$ (Theorem 2.1.6 of \cite{ufp13}, also formalized in this work).
This example demonstrates all the functionality described above in action: implicit elaboration of assumptions throughout all goal and proof statements, automatic definitions for the terms \code{horiz\_pathcomp} and \code{horiz\_pathcomp'} from their constructions via proofs, as well as path induction and calculational reasoning on equalities in the proof of the final lemma.

\section{Discussion and Future Work}

Isabelle/HoTT and its accompanying formalization show that Isabelle's simply typed logical framework infrastructure is feasibly able to provide strong support for modern-day developments of dependent type theory.
However, many improvements are still possible, and future work aims to implement inductive and higher inductive types, as well as to explore how the techniques presented in this paper may be used to implement cubical type theory \cite{ahh18, cchm18} and two-level type theory \cite{acks19, voe13}.

It would be productive to attempt to formalize the notion of a semisimplicial type \cite{her15} in Isabelle/HoTT.
Internalizing the full definition of such an object in homotopy type theory is a well known open problem, with the current state-of-the-art requiring a two-level type theory in order to have a strict equality and natural number type on the outer level \cite{ack16}.
In principle, Isabelle's logical framework can easily provide these.
The main hurdles would again be in implementing enough features on the object logic level, for example to support mutually inductive datatypes.
In this way, the goal of formalizing semisimplicial types could provide further impetus to the development of homotopy type theory in Isabelle.



\bibliography{refs}


\end{document}